\documentclass[12pt]{amsart}
\usepackage{amsmath, amsthm, latexsym, amssymb, amsfonts, epsfig, epsf, xcolor, hyperref}
\usepackage{mathtools}
\usepackage{bm,xcolor,tikz,hyperref}
\usepackage{bm}
\usepackage{blkarray}
\usepackage{booktabs}
\usepackage{enumitem}

\theoremstyle{plain}
\newtheorem{theorem}{Theorem}[section]

\newtheorem{example}[theorem]{Example}

\newcommand{\rmv}[1]{}

\begin{document}
\title{Coding theory package for Macaulay2}
%%%%%%%%%%%%%%%%%%%%%%%
\author{Taylor Ball}
\address[Taylor Ball]{Department of Mathematics\\University of Notre Dame\\ Notre Dame, IN USA}
\email{trball13@gmail.com}
%%%%%%%%%%%%%%%%%%%%%%%
\author{Eduardo Camps}
\address[Eduardo Camps]{Escuela Superior de F\'isica y Matem\'aticas \\ Instituto Polit\'ecnico Nacional\\ Mexico City, Mexico}
\email{camps@esfm.ipn.mx}
%%%%%%%%%%%%%%%%%%%%%%%
\author{Henry Chimal-Dzul}
\address[Henry Chimal-Dzul]{Department of Mathematics and Center of Ring Theory and its Applications\\ Ohio University\\ Athens, OH USA}
\email{hc118813@ohio.edu}
%%%%%%%%%%%%%%%%%%%%%%%
\author{Delio Jaramillo-Velez}
\address[Delio Jaramillo-Velez]{
Departamento de
Matem\'aticas\\
Centro de Investigaci\'on y de Estudios
Avanzados del
IPN\\
Apartado Postal
14--740 \\
07000 Mexico City, D.F.
}

\email{djaramillo@math.cinvestav.mx}
%%%%%%%%%%%%%%%%%%%%%%%
\author{Hiram H. L\'opez}
\address[Hiram H. L\'opez]{Department of Mathematics and Statistics\\ Cleveland State University\\ Cleveland, OH USA}
\email{h.lopezvaldez@csuohio.edu}
\thanks{}
%%%%%%%%%%%%%%%%%%%%%%%
\author{Nathan Nichols}
\address[Nathan Nichols]{School of Mathematics \\ University of Minnesota Twin Cities \\ Minneapolis, MN USA}
\email{nathannichols454@gmail.com}
%%%%%%%%%%%%%%%%%%%%%%%
\author{Matthew Perkins}
\address[Matthew Perkins]{Department of Mathematics\\ Cleveland State University\\ Cleveland, OH USA}
\email{m.r.perkins73@vikes.csuohio.edu}
%%%%%%%%%%%%%%%%%%%%%%%
\author{Ivan Soprunov}
\address[Ivan Soprunov]{Department of Mathematics\\ Cleveland State University\\ Cleveland, OH USA}
\email{i.soprunov@csuohio.edu}
%%%%%%%%%%%%%%%%%%%%%%%
\author{German Vera-Mart\'inez}
\address[German Vera-Mart\'inez]{Escuela Superior de F\'isica y Matem\'aticas \\ Instituto Polit\'ecnico Nacional\\ Mexico City, Mexico}
\email{gveram1100@alumno.ipn.mx}
%%%%%%%%%%%%%%%%%%%%%%%
\author{Gwyn Whieldon}
\address[Gwyn Whieldon]{}
\email{gwyn.whieldon@gmail.com}
%%%%%%%%%%%%%%%%%%%%%%%


\begin{abstract}
In this Macaulay2 \cite{M2} package we define an object called {\it linear code}. We implement functions that compute basic parameters and objects associated with a linear code, such as generator and parity check matrices, the dual code, length, dimension, and minimum distance, among others. We define an object {\it evaluation code}, a construction which allows to study linear codes using tools of
algebraic geometry and commutative algebra. We implement functions to generate important families of linear codes such as Hamming codes, cyclic codes, Reed--Solomon codes, Reed--Muller codes, Cartesian codes, monomial--Cartesian codes, and toric codes. In addition, we define functions for the syndrome decoding algorithm and locally recoverable code construction, which are important tools in applications of linear codes. The package \textit{CodingTheory.m2} is available at \url{https://github.com/Macaulay2/Workshop-2020-Cleveland/tree/CodingTheory/CodingTheory}
\end{abstract}

\maketitle
\markleft{Ball, Camps, Chimal-Dzul, Jaramillo-Velez, L\'opez, Nichols, Perkins, Soprunov, Vera, Whieldon}
%%%%%%%%%%%%%%%%%%%%%%%%%%%%%%%%%%%%%%%%%%%%%%%%%%%%%%%%%%%%%%%%%%%%%%%%%%%%%%%%%%%%%%%%%%%%%%%%%%%%%%%%%%%%%%%%%%%%%%%%%%%%%%%%%%%%%%%%%%%%%%%%%%%%%%%%%%%%%%%%%%%%%%%%%%%%%%
\section{Introduction}
Coding theory has been extensively studied since 1949, when Claude Shannon proved in his seminal paper \cite{Shannon} that linear codes can be used to reliably transmit information from a single source to a single receiver through a noisy channel. Since then, coding theory has found many important engineering applications. For example, coding theory has been used in designing reliable data storage systems, radio communication protocols, and in the emerging field of quantum computers. Coding theory has close ties with many areas in mathematics including linear algebra, commutative algebra, algebraic geometry, and combinatorics. 

In this note we introduce a new package written for Macaulay2 \cite{M2} called \textit{CodingTheory}. The goal of this package is to provide a range of functions for constructing  linear and evaluation codes over finite fields, and for computing some of their main properties. To this aim, we define two objects, namely \textit{linear code} and \textit{evaluation code}. The package also includes implementation of functions for generating important families of linear codes like Hamming codes, cyclic codes, Reed-Solomon codes, Reed-Muller codes, Cartesian codes, monomial-Cartesian codes and toric codes. It also has functions for the syndrome decoding algorithm and locally recoverable codes. 

The organization of this note is as follows. In Section~\ref{def} we describe different ways to define a linear code over a finite field using the \textit{CodingTheory} package. In Section~\ref{pro} we show how to compute the main parameters of a linear code: length, dimension, and minimum distance. We also illustrate how to compute some of the main algebraic objects associated with linear codes like generator and parity check matrices, dual codes, etc. In Section~\ref{eva} we give a brief introduction to evaluation codes and describe some functions implemented to study these objects. In Section~\ref{families} we explain how to create some of the most studied families of linear codes, including Hamming codes, cyclic codes, Reed-Solomon codes, and Reed-Muller codes. Finally, we give instructions on how to create locally recoverable codes.

In this paper we do not attempt to fully explain every function distributed in this package.
%It is important to mention that the functions are not fully explained in this paper. 
For a detailed explanation of all functions in the package, we refer to the Macaulay2 help page which can be accessed by
running 
%installPackage("CodingTheory", RemakeAllDocumentation=>true)
\begin{verbatim}
viewHelp CodingTheory
\end{verbatim}

More information about basics of coding theory can be found in \cite{huf-pless,MacWilliams-Sloane,van-lint}. Constructions of codes using commutative algebra as evaluation codes can be seen in
\cite{carvalho4,GLS,Ha1,LSc,MPV,MPV2,algcodes,renteria-tapia-ca2, Ru, SoSo, So1}. Excellent references for theory of vanishing ideals and their properties are \cite{CLO,monalg}.

\section{Defining linear codes}\label{def}
Let $\mathbb{F}_q$ be a finite field with $q$ elements. Mathematically, a {\it linear code} is defined as a vector subspace $C\subseteq \mathbb{F}_q^n.$ For Macaulay2 (M2), a linear code is a submodule of $\mathbb{F}_q^n.$ Assume $q=p^r,$ where $p$ is a prime number and $r$ a positive integer. By definition, the dual code $C^\perp$ is the orthogonal complement of $C$ in $\mathbb{F}_q^n$ with respect to the standard inner product. One can define $C$ by specifying a list $L$ of elements of $\mathbb{F}_q^n$ that span $C$ or by giving a
a {\it generator matrix} $G$ whose rows form a basis of $C$. Alternatively, one can specify a list $L_{H}$ of elements of $\mathbb{F}_q^n$ that span the dual  code $C^\perp$ or a {\it parity check matrix} $H$ whose columns form a basis of the dual code $C^\perp$.
Below are the commands for the constructor linearCode to construct these equivalent instances of the LinearCode type:
\begin{itemize}
\item linearCode($\mathbb{F}_q$,$L$)
%\item linearCode($\mathbb{F}_q^n$, $L$)
%\item linearCode($\mathbb{F}_q^n$, $L_H$, ParityCheck $=>$ true)
%\item linearCode($\mathbb{F}_q$,$L_{H}$, ParityCheck $=>$ true)
\item linearCode($\mathbb{F}_q$,$n$,$L$)
\item linearCode($G$)
\item linearCode($\mathbb{F}_q$,$L_H$, ParityCheck $=>$ true)
\item linearCode($H$, ParityCheck $=>$ true)
\item linearCode($p$,$r$,$n$,$L$)
\item linearCode($p$,$r$,$n$,$L_H$, ParityCheck $=>$ true)
\end{itemize}

Now, here is a more specific example of how to construct a simple linear code: 

\begin{example}
$\,$
\begin{verbatim}
i2 : F = GF 4;
i3 : L = {{1,1,0,0},{0,0,1,1}};
i4 : C = linearCode(F,L)
o4 = Code with Generator Matrix: | 1 1 0 0 |
                                 | 0 0 1 1 |
o4 : LinearCode
\end{verbatim}
\end{example}

One way to refer to a primitive element of a finite field is by specifying a symbol using the Variable option of the constructor GF.

\begin{example}
$\,$
\begin{verbatim}
i2 : F = GF(9,Variable => a);
i3 : LH = {{1,0,a,0,0},{0,a,a+1,1,0},{1,1,1,a,0}};
i4 : C = linearCode(F,LH,ParityCheck => true)
o4 = Code with Generator Matrix: | a-1 0 a+1 1 0 |
                                 | 0   0 0   0 1 |
o4 : LinearCode
\end{verbatim}
\end{example}
To construct a linear code from a matrix, it is necessary to correctly specify the underlying field. This can be done by passing a field to the matrix constructor.
\begin{example}
$\,$
\begin{verbatim}
i2 : F = GF 4;

i3 : M = matrix(F, {{1,0,1,0},{0,1,1,1}});

             2       4
o3 : Matrix F  <--- F

i4 : C = linearCode(M)

o4 = Code with Generator Matrix: | 1 0 1 0 |
                                 | 0 1 1 1 |

o4 : LinearCode

\end{verbatim}
\end{example}

\section{Basic parameters linear codes}~\label{pro}

The {\it dimension\/} and the {\it length\/} are two of the basic parameters of a code $C\subseteq \mathbb{F}_q^n$. They are defined as the subspace dimension $\dim_{\mathbb{F}_q} C$ and the ambient space dimension $n$, respectively. A third basic parameter is the {\it minimum weight}, which is given by 
\[
\min\{\|c\|
\colon  c \in C, c \neq 0\},
\]
where $\| c \|$ is the number of non-zero entries of $c$. The {\it rate} of $C$ is defined as the rational number $k/n.$ %The dual of $C$ is defined as the orthogonal complement of $C$ in $\mathbb{F}^n_q$ with respect to the usual dot product in $\mathbb{F}^n_q$.
 Some of the functions that can be used in M2 to compute basic parameters and algebraic objects associated with linear codes are the following: \newline
\begin{minipage}[t]{0.33\textwidth}
\begin{itemize}[leftmargin=1 cm]
\item C.GeneratorMatrix
\item C.Generators
\item C.ParityCheckMatrix
\item C.AmbientModule
\item alphabet C
\end{itemize}
\end{minipage}
\begin{minipage}[t]{0.33\textwidth}
\begin{itemize}
\item field C
\item informationRate C
\item ambientSpace C
\item length C
\item dim C
\end{itemize}
\end{minipage}
\begin{minipage}[t]{0.33\textwidth}
\begin{itemize}
\item minimumWeight C
\item codewords C
\item dualCode C
\item shorten(C, List)
\item ==
\end{itemize}
\end{minipage}

\begin{example}
$\,$
\begin{verbatim}
i2: F = GF 4;
i3 : L = {{1,1,0,0},{0,0,1,1}};
i4 : C = linearCode(F,L)
o4 = Code with Generator Matrix: | 1 1 0 0 |
                                 | 0 0 1 1 |
o4 : LinearCode
i5 : length C
o5 = 4
i6 : dim C
o6 = 2
i7 : informationRate C
     1
o7 = -
     2
o7 : QQ
i8 : ambientSpace C
      4
o8 = F
o8 : F-module, free
i9 : alphabet C
o9 = {0, a, a + 1, 1}
o9 : List
i10 : minimumWeight C
o10 = 2
i11 : dualCode C
o11 = Code with Generator Matrix: | 1 1 0 0 |
                                  | 0 0 1 1 |

o11 : LinearCode
\end{verbatim}
\end{example}

\section{Evaluation codes}\label{eva}
Let $\mathcal{X}=\left\{{\bf a}_1,\ldots,{\bf a}_n\right\}$ be a subset of an $m$-dimensional space $\mathbb{F}_q^{m}$.
Consider a finite dimensional subspace $S\subset {\mathbb{F}_q}[X_1,\ldots,X_m]$  of the $m$-variate polynomial ring over $\mathbb{F}_q$.
%and $S_{\leq d}$ the ${\mathbb{F}_q}$-vector space of all polynomials in $S$ of degree at most $d.$ 

The {\it evaluation map\/}
\begin{equation*}\label{functionevaluation}
{\rm ev}_S\colon S\longrightarrow {\mathbb{F}_q}^{|\mathcal{X}|},\ \ \ \ \ 
f\mapsto \left(f({\bf a}_1),\ldots,f({\bf a}_n)\right),
\end{equation*}
defines a linear map of ${\mathbb{F}_q}$-vector spaces. The image of ${\rm ev}_S$ in ${\mathbb{F}_q}^{|\mathcal{X}|}$, denoted by $C_{\mathcal{X}}(S)$, %defines a ${\mathbb{F}_q}$-vector subspace. We refer to $C_{\mathcal{X}}(d)$ 
is the {\it evaluation code\/} on the set $\mathcal{X}$ corresponding to $S$. The {\it vanishing ideal} of $\mathcal{X}$, denoted by $I(\mathcal{X})$, is the ideal in $S$ of all polynomials that vanish on  $\mathcal{X}.$ A key observation that allows the use of commutative algebra in studying evaluation codes is that the kernel of the evaluation map ${\rm ev}_S$ is precisely $S\cap I(\mathcal{X})$.

An evaluation code $C_{\mathcal{X}}(S)$ is defined in M2 as a separate type because there are more objects associated with it than with a linear code. For instance, the vanishing ideal associated to the set $\mathcal{X}$ plays an important role when finding and estimating parameters of the code, so it is convenient to be able to access it. Given an evaluation code $C$ in M2, the object C.linearCode is a linear code in M2. The command $\text{evaluationCode}(\mathbb{F}_q,\mathcal{X},L)$ defines an evaluation code where $\mathcal{X}$ is a list of elements in $\mathbb{F}_q^{m}$ and $L$ is a list of polynomials that span $S$. In the case when polynomials in $L$ are monomials, one may give the matrix of exponent vectors instead of $L$. 

There are many construction of evaluation codes for specific choices of the set $\mathcal{X}$ and the subspace $S$. These include
Reed-Muller codes, Cartesian and monomial Cartesian codes, toric codes, and evaluation codes from graphs. We refer to 
\cite{carvalho4, Ha1, LSc, lopez-villa, MPV, algcodes, Ru, SoSo} for details on how these codes are defined and what properties they have from coding theory, commutative algebra, and algebraic geometry perspectives. Some functions defined in this package for various constructions of evaluation codes and associated algebraic objects are the following:

\medskip

\begin{minipage}[t]{0.52\textwidth}
\begin{itemize}[leftmargin=0.55 cm]
\item evaluationCode($\mathbb{F}_q$,List,List)
\item toricCode($\mathbb{F}_q$,Integer matrix)
\item cartesianCode($\mathbb{F}_q$,List,List)
\item orderCode($\mathbb{F}_q$,List,List,ZZ)
\item evCodeGraph($\mathbb{F}_q$,Incident Matrix,integer)
\end{itemize}
\end{minipage}
\begin{minipage}[t]{0.45\textwidth}
\begin{itemize}[leftmargin=1 cm]
\item vNumber(Ideal)
\item footPrint(Integer,Integer,Ideal)
\item hYpFunction(Integer,Integer,Ideal)
\item gMdFunction(Integer,Integer,Ideal)
\item vasFunction(Integer,Integer,Ideal)
\end{itemize}
\end{minipage}

\medskip

The mathematical definitions of the vNumber, the footprint function, the hyp function, the generalized footprint function and the Vasconcelos function can be found in~\cite{CSTVV}. The following example shows how to construct an evaluation code.

\begin{example}
$\,$
\begin{verbatim}
i2 : F=GF(4); R=F[x,y,z];
i4 : P={{0,0,0},{1,0,0},{0,1,0},{0,0,1},{1,1,1},{a,a,a}};
i5 : S={x+y+z,a+y*z^2,z^2,x+y+z+z^2};
i6 : C=evaluationCode(F,P,S)
o6 = Code with Generator Matrix: | 0 1 1 1 1   a   |
                                 | a a a a a+1 a+1 |
                                 | 0 0 0 1 1   a+1 |
                                 | 0 1 1 0 0   1   |
o6 : EvaluationCode
i7 : length C.LinearCode
o7 = 6
i8 : dim C.LinearCode
o8 = 3
i9 : C.Points
o9 = {{0, 0, 0}, {1, 0, 0}, {0, 1, 0}, {0, 0, 1}, {1, 1, 1}, {a, a, a}}
o9 : List
i10 : C.VanishingIdeal;
o10 = Ideal of R
\end{verbatim}
\end{example}
\section{Families of linear codes}\label{families}
We continue with the same notation: $n$ represents the length of the code, $k$ the dimension and $q$ the size of the field.
Some families of linear codes that have been implemented in this package are the following:

\medskip

\begin{minipage}[t]{0.5\textwidth}
\begin{itemize}[leftmargin=1 cm]
\item HammingCode($q$,integer)
\item randLDPC($n$, $k$, integer, integer)
\item cyclicCode($\mathbb{F}_q$ ,polynomial, $n$)
\item quasiCyclicCode($\mathbb{F}_q$,list)
\item RSCode($\mathbb{F}_q$,List,integer)
\item RMCode($q$,integer,integer)
\end{itemize}
\end{minipage}
\begin{minipage}[t]{0.5\textwidth}
\begin{itemize}[leftmargin=1 cm]
\item zeroCode($\mathbb{F}_q$,$n$)
\item universeCode($\mathbb{F}_q$,$n$)
\item repetitionCode($\mathbb{F}_q$,$n$)
\item zeroSumCode($\mathbb{F}_q$,$n$)
\item random($\mathbb{F}_q$, $n$, $k$)

\end{itemize}
\end{minipage}

\medskip

Mathematical definitions of the above families can be found in \cite{huf-pless,MacWilliams-Sloane,van-lint}.

\begin{example}
$\,$
\begin{verbatim}
i2 : C = HammingCode(2,3)
o2 = Code with Generator Matrix: | 1 1 1 1 0 0 0 |
                                 | 0 1 1 0 1 0 0 |
                                 | 1 0 1 0 0 1 0 |
                                 | 1 1 0 0 0 0 1 |
o2 : LinearCode
i3 : F=GF(5); R=F[x]; g=x-1; C=cyclicCode(F,g,6)
Cyclic Code
o6 = Code with Generator Matrix: |-1  1  0  0  0  0 |
                                 | 0 -1  1  0  0  0 |
                                 | 0  0 -1  1  0  0 |
                                 | 0  0  0 -1  1  0 |
                                 | 0  0  0  0 -1  1 |
o6 : LinearCode
i7 : F = GF(5);
i8 : L = apply(toList(1..2),j-> apply(toList(1..4),i-> random(F)))
o8 = {{0, -2, 2, -1}, {-2, 1, 0, -1}}
o8 : List
i9 : C=quasiCyclicCode(L)
o9 = Code with Generator Matrix: |  0 -2  2 -1 |
                                 | -1  0 -2  2 |
                                 |  2 -1  0 -2 |
                                 | -2  2 -1  0 |
                                 | -2  1  0 -1 |
                                 | -1 -2  1  0 |
                                 |  0 -1 -2  1 |
                                 |  1  0 -1 -2 |
o9 : LinearCode
i7 : C=RSCode(GF 5,{1,2,3},3)
o7 = Code with Generator Matrix: | 1  1  1 |
                                 | 1  2 -2 |
                                 | 1 -1 -1 |
o7 : EvaluationCode
\end{verbatim}
\end{example}
\section{Applications of linear codes}
A basic application of a linear code is {\it decoding}, which is used for reliable transmission of information trough a noisy channel. In a few words the idea is the following. Take a vector $c\in C.$ Change the value of some of the entries of $c$ to obtain a new vector $v$. Decoding the vector $v$ means to recover the vector $c$ when only $v$ and $C$ are given. Detailed treatment of decoding algorithms can be found in \cite{huf-pless}. Another, more recent application of linear codes for found in distributed and cloud storage systems. The idea is to use {\it locally recoverable codes}, which are linear codes with the property that every entry can be recovered from a few other entries. For more information on locally recoverable codes see~\cite{TamoBarg}.

Some of the most important functions from this package that can be used for applications of coding theory are the following:
\begin{itemize}
\item syndromeDecode(C,$v$,minimumWeight(C))
\item LocallyRecoverableCode(List,List,a polynomial)
\end{itemize}

Here is a small example. 

\begin{example}
$\,$
\begin{verbatim}
i2 : C = HammingCode(2,3);
i3 : msg = matrix {{1,0,1,0}};
i4 : v = msg*(C.GeneratorMatrix)
o4 = | 0 1 0 1 0 1 0 |
i5 : err = matrix take(random entries basis source v, 1)
o5 = | 0 0 0 0 1 0 0 |
i6 : received = transpose(transpose (v+err))
o6 = | 0 1 0 1 1 1 0 |
i7 : transpose syndromeDecode(C, transpose recieved, 3)
o7 = | 0 1 0 1 0 1 0 |
\end{verbatim}
\end{example}

\section*{Acknowledgments}
We thank Federico Galetto, Courtney Gibbons, Hiram L\'opez, and Branden Stone for organizing the Macualay2 workshop at Cleveland State University, where this collaboration started. We want to give a special thanks to Branden Stone for helping us to develop the package during the workshop.

\end{document}